# An Evolution Model of Complex Systems with Simultaneous Cooperation and Competition

Xiu-Lian Xu, Chun-Hua Fu, Hui Chang, Da-Ren He*

*College of Physics Science and Technology, Yangzhou University, Yangzhou 225002, China*

(* Corresponding author: darendo10@yahoo.com.cn)

**Abstract**

Systems with simultaneous cooperation and competition among the elements are ubiquitous. In spite of their practical importance, knowledge on the evolution mechanism of this class of complex system is still very limit. In this work, by conducting extensive empirical survey to a large number of cooperation-competition systems which cover wide categories and contain the information of network topology, cooperation-competition gain, and the evolution time, we try to get some insights to the universal mechanism of their evolutions. Empirical investigations show that the distributions of the cooperation-competition gain interpolates between power law function and exponential function. Particularly, we found that the cooperation-competition systems with longer evolution durations tend to have more heterogeneous distributions of the cooperation-competition gain. Such an empirical observation can be well explained by an analytic model in which the evolution of the systems are mainly controlled by the Matthew effect, and the marginal heterogeneity of the initial distribution is amplified by the Matthew effect with similar speed in spite of the diversity of the investigated systems.
**Key Words:** cooperation-competition; cooperation sharing; heterogeneity
**PACS:** *89.75.-k; 87.23.Kg; 89.75.Da*

## 1. Introduction

Simultaneous cooperation and competition is the common feature for a number of complex systems, including the commercial [1-2], ecological [3-5], social [6-8], biochemical [8], economical [8-10] and some technological systems [11-18]. Revealing the universal rules and corresponding mechanisms controlling the evolution of such cooperation-competition (C-C) systems attracts much attention from the scientists in physics. In literature, several models have been proposed to explain the empirically observed statistical properties of this kind of C-C systems [4-10]. For example, in Ref. [4], to explain the commonly observed exponential truncation in the degree distribution of the mutualistic ecological networks, Guimaraes Jr. and coworkers developed a model by introducing the "build-up" mechanism. The resulted degree distribution based on their model interpolates between the power law and exponential decay, suggesting that some ecological factors confined the appearance of large hubs (or the accretion of the degree distribution heterogeneity) [4]. In Ref. [10], by considering the Matthew effect (rich gets richer) in the growth rate of the so-called "competition gain", the author developed an evolution model which can successfully reproduce the observation of wealth condensation in economic systems.

The success of the above models demonstrates that it is important to simultaneously consider the cooperation and competition in modeling the evolution of the cooperation-competition systems. However, all these previous work focused on the specific system(s). One interesting question is whether the evolution mechanism derived based on these specific systems is also applicable to other cooperation-competition systems, or equivalently, can we derived a certain universal rule which controls the evolution of the whole class of cooperation-competition systems? Answering to this question is very difficult since it requires large number of empirical data covering wide categories. To our knowledge, such kind of work is still lacking up to now.

In this work, as an effort to get some insights to the above question, we try to collect a number of cooperation-competition systems with large diversity. The collected systems include not only the data of network topology but also the data of cooperation-competition gain (in this work, we call it cooperation sharing (CS)). More importantly, many of the systems also contain the data of time during of the system evolution, which makes it possible to directly extract the information of the evolution dynamics by empirical survey. Our empirical results show that the distributions of the cooperation sharing for all the studied systems interpolate between power law function and exponential function. Particularly, the degree of heterogeneity of the distribution for each system is strongly correlated with the time duration of the system evolution. The systems with longer evolution durations tend to have more heterogeneous distributions of the cooperation sharing. Such an empirical observation can be well explained by an analytic model in which the evolution of the systems are mainly controlled by the Matthew effect, and the marginal heterogeneities of the initial distributions are modified by the Matthew effect with the similar speed in spite of the large diversity of the investigated systems. Our results revealed a general feature for this kind of cooperation-competition systems, contributing to the understanding of the complex systems with simultaneous cooperation and competition.

The rest parts of the article are organized as follows. In section 2 we will introduce the collections of the cooperation-competition systems investigated in this work, and some empirical results are provided. Based on the empirical results, an evolution picture of the cooperation-competition systems is proposed. In section 3, an analytic discussion on the model will be presented. In section 4 we will give the comparisons between the analytic results derived based on our model and the empirical observations. In the last section, a summary and some discussions will be presented.

**2. Data collection and empirical investigations**

In the cooperation-competition systems we studied, elements cooperate and compete in some groups [11-18]. The groups, which are the platform of the cooperation and competition among the elements, can be places, organizations, events, or activities. The elements are the participants of the cooperation-competition, and they can be human beings or abiotic things that are subjected to influences of human beings. The same element may be involved in a number of groups simultaneously. For example, to describe the cooperation and competition among the movie actors, the movie actors who are casted in a certain movie are considered as elements of the C-C system, while the movies are considered as groups [18]. Similarly, when studying the evolution of the human languages, the languages, which coexist in a geographical region, can be considered as elements, and the geographical regions can be considered as groups. During the long time evolution, the languages compete for being used by more people [18]. In each group of a C-C system, some elements make concerted effort to accomplish a task, which often create a type of product. The product may induce several kinds of resources. The elements, when they cooperate, also compete for a larger piece of the resources. For example, some Hollywood actors work together to produce a movie, which should bring box office income (a countable resource) and famousness (an uncountable resource). Meanwhile, they compete for playing more important roles in the movie, which usually means higher salaries and famousness. In another example, some languages cooperate for creating an "oral and written communication" of some different kinds of people in a group (region). The produced resource of the cooperation is the populations who live together in the region and use some of the languages. In a historical period, the languages compete for being used by more people (that is why some languages got extincted, but some other languages have been used by more and more people and spread to more and more geographical regions). The number of people who use a certain language can be regarded as a countable resource, which the element (the language) shares.

In this work, we use $h_i$ to denote the group number in which the $i$th element takes part. It can be expressed as $h_i = \sum_j b_{ij}$ with $b_{ij} = 1$ if element $i$ is involved in group $j$ and $b_{ij} = 0$ otherwise. The so-called "group size", $T_j$, denotes the number of the elements which take part in group $j$. $T_j$ can be expressed as $T_j = \sum_i b_{ij}$ [11,12]. In Ref. [13], Fu et al. defined the CS as the part of a countable resource which the element shares. Considering that the competition intensity should depend on the group size, in this work the total CS of element $i$ in group $l$ is defined as $W_{li}=T_l z_{il}$, where the $z_{il}$ denotes the countable resource shared by element $i$ in group $l$. The normalized total cooperation sharing (NTCS) of the $i$th element, $\omega_i$, can be defined as $\omega_i = (\sum_l T_l z_{il}) / \sum_j [(\sum_l T_l z_{jl})]$ [14,16,18].

In constructing the real world C-C systems, the data were collected based on the following rules: (1) All the elements can be unambiguously assigned to the related groups; (2) The information of the amount of CS shared by the elements in each group is available; (3) The evolution duration of each group (this concept will be introduced bellow) is available. We totally constructed more than 60 systems, among which only 12 systems have all the reliable information. For example, Hollywood movie actor cooperation may be a famous C-C system. Unfortunately, it is difficult to get the CS data. Therefore, the 12 systems can be considered to be randomly selected and reasonably cover the essential features of the C-C systems. The 12 C-C systems are listed in Table 1, which include 1) Chinese university matriculation (CUM), 2) the 2004 Athens Olympic Game (OG), 3) notebook PC selling at Taobao website (PCST), 4) information technique (IT) product selling (ITPS), 5) journal impact factor system (JIF), 6) author academic level system (AAL), 7) the 200 richest Chinese magnates in 2004 (RCM), 8) the USA county population system in 1900 (UCO19), 9) the USA county population in 2000 (UCO20), 10) Beijing restaurant system (BR), 11) mixed drink (MD) (e.g., cocktails) system, and 12) the world language distribution system (WLD). More details on the data collections and constructions of these C-C systems can be found in Ref. [18]. In most cases, C-C among the elements occurs right after the birth of the groups; therefore, we can define the evolution duration as the time duration between the group birth and the group termination or data collection. We define the "group birth" as the time when the first element joins the group, and the "group termination" as the time when the cooperation task is accomplished and the elements disband.

The description of C-C dynamics certainly will be in a logarithmic time scale because the evolution durations of different systems must be very different and cover many orders of magnitudes. We emphasize that for the current purpose, it is sufficient that the data of the evolution duration are reliable within one order of magnitude, although the exact evolution duration for all the groups may be obtained. In some systems there are many groups, and the groups show different evolution durations. It is meaningless to list the evolution durations for all the groups. As listed in Table 1, we use the longest and the shortest evolution durations, $\tau_{max}$ and $\tau_{min}$, among the groups of the C-C system, as well as the average of them in the current work, which are sufficient to capture the order of the magnitude and the error range of the evolution durations. More detailed discussions were given in Ref. [18].

In Ref. [18] we presented the details of the definitions and interpretations for the 12 real world C-C systems. The NTCS distributions for all these systems were also reported. The results showed that all the distributions obey the so-called "shifted power law (SPL) functions" which can be expressed as $P(x) \propto (x+\alpha)^{-\gamma}$ [12]. When $\alpha = 0$, the SPL function is reduced to a power law function. When $\alpha \rightarrow 1$, we can prove that the SPL function is reduced to an exponential decay function in the condition that $x$ is normalized (i.e., $0<x_i<1$ and $\sum_{i=1}^{M} x_i = 1$, $M$ denotes the total number of $x$) [18]. Therefore an SPL interpolates between a power law and an exponential decay. The parameter $\alpha$ characterizes the degree of deviation from a power law. We emphasize that in general an SPL is not a power law, and $\gamma$ is not the power law scaling exponent. It is not strange to observe a very large $\gamma$ as will be discussed later. We will show that both the $\gamma$ and $\alpha$ characterize the heterogeneity of the distributions, and that, very possibly, the $\gamma$ and $\alpha$ keep a general correlation which encodes the information of the evolution mechanism of the C-C systems. The definitions of the elements and the groups, the empirical evolution durations, and the parameters of the NTCS distribution function for each of the C-C systems are listed in Table 1.

Table 1. The interpretations and the numbers of groups and elements. Two parameter values, $\gamma$ and $N\alpha$ ($N$ denotes element number) of the NTCS distribution functions $P(x) \propto (x+\alpha)^{-\gamma}$. The longest and the shortest group evolution durations, $\tau_{max}$ and $\tau_{min}$, and the averaged evolution duration $\tau$ are also listed.

| Syst. No. | System | Group | element | CS | group No. | element No. | $\gamma$ | $N\alpha$ | $\tau_{max}$ (year) | $\tau_{min}$ (year) | $\tau$(year) |
|---|---|---|---|---|---|---|---|---|---|---|---|
| 1 | CUM | Batch | University | Mark | 51 | 2277 | $3.7 \times 10^3$ | $2.3 \times 10^3$ | 0.03 | 0.008 | 0.016 |
| 2 | OG | Event | athlete | Score | 133 | 4500 | $3.6 \times 10^3$ | $4.5 \times 10^3$ | $6.9 \times 10^{-4}$ | $3.2 \times 10^{-7}$ | $2.3 \times 10^{-4}$ |
| 3 | PCST | Market | shop | Price | 53 | 4711 | 6.1 | 4.2 | 3 | 3 | 3 |
| 4 | ITPS | Market | manufacturer | Rank | 265 | 2121 | 4.5 | 2.9 | 94 | 5 | 28 |
| 5 | JIF | publication | journal | IF | 1 | 6559 | 3.9 | 1.4 | 186 | 3 | 34 |
| 6 | AAL | publication | author | Level | 1 | 784 | 3.7 | 1.8 | 61 | 8 | 29 |
| 7 | RCM | Market | magnate | Wealth | 1 | 200 | 3.1 | 0.24 | 100 | 5 | 20 |
| 8 | UCO19 | USA | county | population | 1 | 2834 | 3.06 | 0.74 | 124 | 4 | 70 |
| 9 | UCO20 | USA | county | population | 1 | 3139 | 2.2 | 0.32 | 224 | 41 | 160 |
| 10 | BR | Market | restaurant | attention | 688 | 3337 | 2.12 | 0.12 | 159 | 9 | 75 |
| 11 | MD | Drink | ingredient | proportion | 7804 | 1501 | 1.8 | 0.077 | 500 | 100 | 340 |
| 12 | WLD | Country | language | population | 236 | 6142 | 1.6 | 0.0086 | $5 \times 10^3$ | 40 | $2.5 \times 10^3$ |

From Table 1, one can see that the $N\alpha$ ($N$ is the element number) and $\gamma$ basically show a monotonic relation. This implies a possibility that the correlation between the $N\alpha$ and $\gamma$ obeys a general function, which may encode the information of the evolution mechanism of the systems. In the next two sections, we will discuss this further. Note that in this work, we always use the $N\alpha$ when discussing the SPL parameter $\alpha$. As mentioned above, the parameter $\alpha$ characterizes the degree of deviation of the distribution $P(\omega) \propto (\omega+\alpha)^{-\gamma}$ from a power law function. For different systems, it is inappropriate to directly compare the $\alpha$ values since the averaged $\omega$ values can be very different. A more reasonable way is defining a "relative value" of $\alpha$, which can be expressed as $\alpha' = \dfrac{\alpha}{<\omega>} = \dfrac{\alpha}{1/N} = N\alpha$, as used in this work.

Fig. 1 shows the SPL functions best fitting the NTCS distributions for the 12 real world C-C systems. The corresponding parameters $\alpha$ and $\gamma$ of the SPL functions are listed in Table 1. More details of the SPL fitting were provided in Ref. [18]. For clarity, all the fitting lines are translated to have a common point without changing their slopes (i.e., $\gamma$ values of the SPL functions) on the double-logarithmic plane. Note that the lengths of the fitting lines are meaningless, and are used to distinguish two lines with similar slopes.

In Table 1 and Fig. 1, the systems rank in descent order of the $\gamma$ values. Interestingly, one can observe that the averaged evolution duration $\tau$ basically shows a monotonic increase with the deceasing of the $\gamma$ values. Considering that the $\gamma$ values describe the distribution heterogeneities of the CS, the monotonic dependence of the $\gamma$ values on the evolution duration $\tau$ indicates that the cooperation-competition systems with longer evolution durations tend to have more heterogeneous distributions of the CS. This empirical observation strongly supports the following picture of the system evolution: the time duration of the system evolution plays dominant role in controlling the distribution heterogeneity of the CS of the C-C systems, and all these C-C systems follow the similar evolution mechanism, Consequently, the final distribution heterogeneity (described by $\gamma$ [17,19]) only depends on the system evolution duration. Inspired by the previous work in the specific C-C systems [4, 10], we propose that the evolution of the system is dominated by the Matthew effect. Suppose that all the systems start the competition and cooperation at a common time point, the parameter $\gamma$ of each system will get increasingly smaller when it evolves for longer time (i.e., $\tau$ becomes longer). This picture resembles the case of running. The heterogeneity (difference between the athletes) is small in a short distance running, while it becomes larger and larger when the distance becomes longer. As is well known [19], on the $x$-$P(x)$ plane, a line with a larger slope indicates a weaker heterogeneity of the distribution. The most homogeneous distribution shows a line with an infinitely large slope (an ideal vertical line), which means that all the elements possess a common $x$ value. According to the empirical observations that $\gamma$ monotonically depends on $\tau$, we suppose that such a vertical line corresponds to $\tau=0$, as shown in Fig. 1. Thus, in our model we assume that at the beginning ($t=0$), all the systems show even NTCS distributions as shown by the line indicated by $\tau=0$ in Fig. 1. Right after the start of time evolution ($\tau$ becomes larger but is still close to the starting point), the systems suffer a random perturbation so that CS of the elements show marginal differences. The corresponding fitting lines of the NTCS distributions for all the 12 systems should be almost on the same position as denoted by the line indicated by $\tau \approx 0$ in Fig. 1. As the time develops the elements in all the systems simultaneously cooperate and compete. The Matthew effect amplifies the differences of the element CS and the distribution heterogeneity. The system, which has a longer $\tau$, finally shows a fitting line with smaller slope $\gamma$ in Fig. 1, corresponding to a larger CS distribution heterogeneity. In order to show the difference of the four lines with the largest slopes, the lines are shown in the inset of Fig. 1 where the region of x-axis is greatly magnified. We will show that such a simple evolution picture is reasonable and supported by empirical observations.

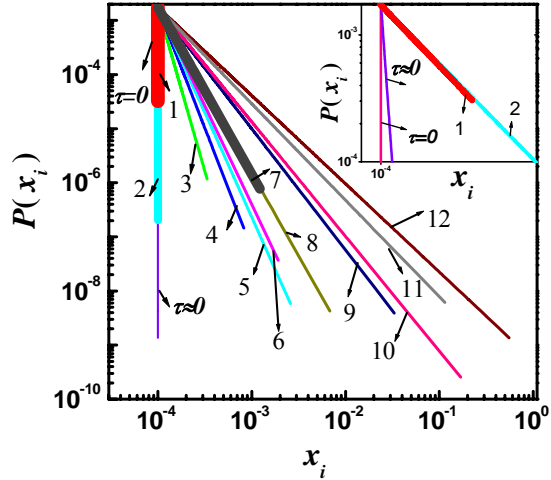

FIG. 1. (color online) The NTCS distribution fitting lines of 12 real world C-C systems. The two lines indicated by $\tau=0$ and $\tau\approx0$ are explained in the text. The inset shows the zoom-in of the region around $x=10^{-4}$.

## 3. Analytic model of the system evolution

The above proposed evolution picture can be easily implemented in an analytic model. In the model, at $\tau=0$ all the elements in any of the systems possess a common NTCS value $\omega_0$. The value can be an arbitrary small number. Without losing generality, in Fig. 1 we choose $\omega_0=10^{-4}$. When $\tau$ becomes larger but still close to zero, the NTCS values of the elements show marginal differences due to certain random perturbations. We sort the $N$ elements according to their NTCS values $\omega_i$ ($i=1,\cdots,N$) by descent order with $i$ being the ranking number, i.e., $\omega_1 \geq \omega_2 \geq \cdots \geq \omega_i \geq \cdots \geq \omega_N$. As the time develops, Matthew effect magnifies the differences between the $\omega$ values so that a large $\omega_i$ gets larger. A simple but appropriate description of the NTCS evolution may be a power law form of $\omega_i$ with a negative exponent which is proportional to the time. Such an assumption of the evolution function is supported by the "B-A model" proposed in Ref. [20]

by Barabasi and Albert, in which the authors analytically derived a power law evolution description of network degree with a negative exponent based on a "rich gets richer" mean field dynamic equation. It is known that the ranking number $i$ should obey a similar power law [19]. Consequently, the NTCS evolution function can be written as

$$W_i(t) = \omega_0 \left(\frac{i}{N}\right)^{-t/c} - \omega_0 \tag{1}$$

where $W_i(t)$ denotes the net CS at time $t$. The normalized form of $W_i(t)$ is given by

$$\omega_i = \frac{W_i}{\sum W_i} \approx \frac{W_i}{\int_{1/N}^{1}[N\omega_0 x^{-t/c}]dx - N\omega_0}. \tag{2}$$

In the next section we will show that although the value of the constant $c$ in Eqs. 1 and 2 does not influence the main results, the value of 100 years is chosen in this work (i.e., $c=c_s=100$ years), with which the analytical results can best reproduce the empirical data. When $\tau/c_s < 1$, from Eqs.1 and 2 we get

$$\frac{i}{N} = \left[\frac{\tau N}{c_s - \tau}\left(\omega_i + \frac{c_s - \tau}{\tau N}\right)\right]^{-c_s/\tau}. \tag{3}$$

where $\tau$ is the time when the system reaches the final state. The probability for finding an element whose NTCS is smaller than $\omega_i$ is given by

$$P(\omega_i' < \omega_i) = \frac{N-i}{N} = 1 - \left[\frac{\tau N}{c_s - \tau}\left(\omega_i + \frac{c_s - \tau}{\tau N}\right)\right]^{-c_s/\tau}. \tag{4}$$

Therefore,

$$P(\omega_i) = \frac{\partial P(\omega_i' < \omega_i)}{\partial \omega_i} = \frac{c_s}{\tau}\left(\frac{\tau N}{c_s - \tau}\right)^{-c_s/\tau}\left(\omega_i + \frac{c_s - \tau}{\tau N}\right)^{-(1+c_s/\tau)} \propto (\omega_i + \alpha)^{-\gamma}. \tag{5}$$

Since $\tau/c_s < 1$, we have $\gamma(\tau)=(1+c_s/\tau)>2$. As can be seen in Table 1, 10 of the 12 real world systems (No.1-No.10) belong to this type [18]. For these systems, one gets

$$\gamma(\tau) - 1 = c_s/\tau \propto \tau^{-1}, \tag{6}$$

$$N\alpha(\tau) + 1 = c_s/\tau \propto \tau^{-1} \tag{7}$$

and

$$\gamma(\tau) = N\alpha(\tau) + 2. \tag{8}$$

Similarly, when $\tau/c_s = 1$, we get

$$P(\omega_i) \propto (\omega_i + \frac{1}{N(\ln N - 1)})^{-(1+c_s/\tau)} = (\omega_i + \alpha)^{-\gamma}. \tag{9}$$

For this kind of systems, the analytic expression $\gamma(\tau)-1 \propto \tau^{-1}$ is still valid, whereas the relation $\gamma(\tau) = N\alpha(\tau) + 2$ is invalidated since $\alpha$ does not relate to $\tau$.

When $\tau/c_s > 1$, we obtain a similar form as

$$P(\omega_i) \propto (\omega_i + \frac{\tau - c_s}{c_s N^{\tau/c_s} - N\tau})^{-(1+c_s/\tau)} = (\omega_i + \alpha)^{-\gamma}. \tag{10}$$

In this case, we have $\gamma(\tau)=(1+c_s/\tau)<2$. In Table 1 one can see that only 2 real world systems, (No.11) and (No.12), are of this type, and that the systems show a very long $\tau$ value [18]. Clearly, the analytic conclusion $\gamma(\tau)-1 \propto \tau^{-1}$ is still effective. The systems should show very small $\alpha$ values, as shown in Table 1.

## 4. Comparison between analytic and empirical results

The analytic expression $\gamma(\tau)-1 = c_s/\tau \propto \tau^{-1}$ is the only common conclusion for all the three cases discussed in the last section. Fig. 2 shows the comparison of this rule with the empirical data. In the figure the vertical coordinates of the solid circles denote the empirical data of $\tau$. The vertical coordinates of upper and lower positions of the error bars are drawn according to the empirically obtained longest and the shortest evolution durations, $\tau_{max}$ and $\tau_{min}$, respectively. The solid line is the least-square fitting of all the data. The slope of the fitting line, -1.11, is very close to the analytic value (-1.0). Different $c_s$ values can shift the fitting line upward or downward without changing its slope. When we chose the $c_s$ value of 100 years as mentioned in the last section, the line representing the analytic results will be very close to the fitting line of the empirical data.

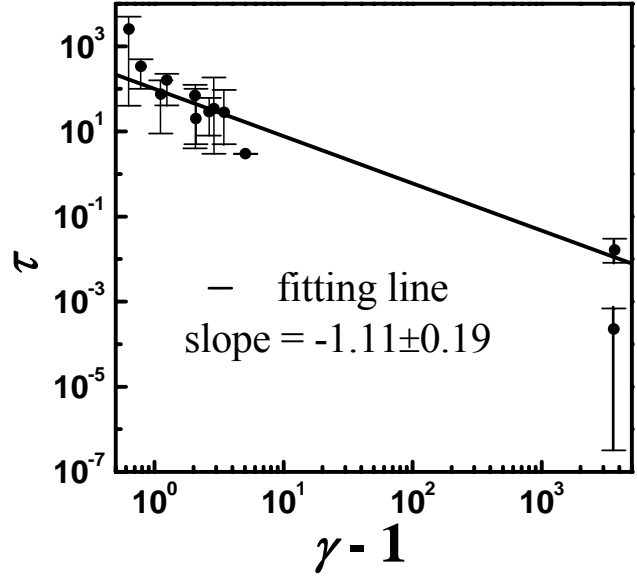

FIG. 2. The empirical relation between $\tau$ and $\gamma$.

From Table 1 one can see that 10 of the 12 real world systems (No.1-No.10) show $\tau/c_s < 1$ and $\gamma(\tau)=(1+c_s/\tau)>2$ [18]. For these systems the analytical conclusion gives $N\alpha(\tau)+1=c_s/\tau \propto \tau^{-1}$ and $\gamma(\tau) = N\alpha(\tau)+2$. Fig. 3 and 4 show the comparison between the theoretical and the empirical results. Similarly, in Fig. 3 the vertical coordinates of the solid circles denote the empirical data of $\tau$. The vertical coordinates of the upper and lower positions of the error bars are drawn according to $\tau_{max}$ and $\tau_{min}$, respectively. The solid line is the least-square fitting of the data. The slope of the fitting line (-1.10) is very close to the analytic value (-1.0). If we chose the $c_s$ value of 100 years, again the line representing the analytic results will be almost identical to the fitting line of the empirical data. In Fig. 4 the crosses denote the empirical data of the 10 systems. The solid line is drawn based on the analytic results, i.e., $N\alpha(\tau)+2=\gamma(\tau)$. Obviously, all the empirical data are in good agreement with the analytic results.

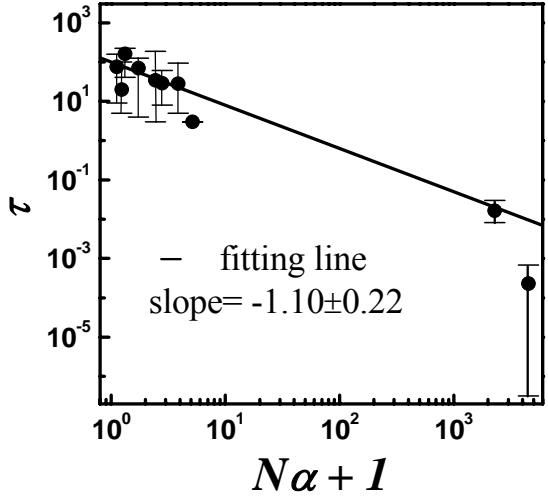 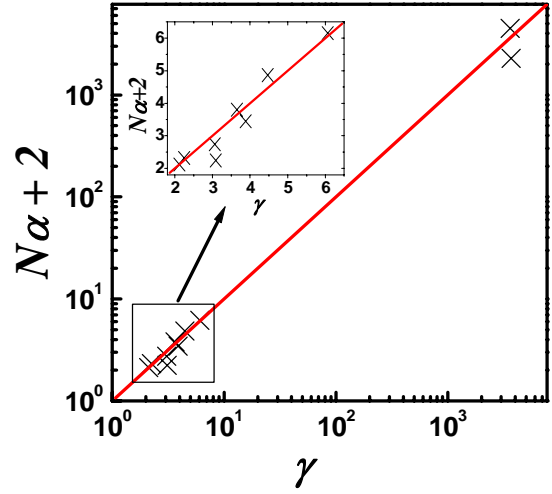

FIG. 3. The empirical relation between $\tau$ and $N\alpha$.　　FIG. 4. The empirical relation between $\gamma$ and $N\alpha$.

## 5. Summary and discussions

This article shows that the CS distributions in all the C-C systems interpolate between the power law function and exponential function, which can be described by the SPL function. Both the distribution parameters of SPL function, $\alpha$ and $\gamma$, signify the heterogeneities of the distributions. Based on the empirical investigations obtained for 12 real world C-C systems, we proposed a very simple C-C evolution model where Matthew effect dominates so that different systems show different heterogeneities of CS distributions only due to the different evolution durations. Based on this model we analytically obtained the correlation functions between the $\alpha$, $\gamma$ and the evolution duration $\tau$, which are in good agreement with the empirical observations. The relationship between the $\alpha$ and $\gamma$, i.e., $\gamma(\tau) = N\alpha(\tau) + 2$, is also in agreement with our previous investigation obtained in a very different way [17], which may suggest that the relationship is general and captures the essence of the evolution dynamics of the C-C systems.

People may raise questions about the model idea proposed above. For example, one may feel hard to accept that Matthew effect is the only dominating factor in the evolution of the C-C systems. It is natural to also consider the opposite factors which tend to make the CS distributions even. To investigate the effects of such even factors, we tried to add random perturbations, which tend to destroy the distribution heterogeneities therefore mimics the effects of the even factors, during the evolution of the systems. The resulted CS distribution shows enhancement of the "middle class" elements, which obviously deviates from empirically observed SPL distributions (data not shown). In comparison, all the empirical investigations on the 12 real world C-C systems support the Matthew effect dominated evolution model proposed in the current work.

In the current model, the even factor is introduced only at the beginning of the competition. This resembles the situations in Chinese history during which many dynasties alternated. Quite often, when an old dynasty suffered perdition due to too large wealth heterogeneity, a new dynasty emerged; this destroyed almost all the heterogeneities. The randomly arising heterogeneity then was magnified in the new dynasty. Differently, a modern government knows well that, in order to keep the society stable, it is very important to maintain the wealth heterogeneity below a threshold. Therefore a modern government often adopts some policies to shrink the wealth heterogeneity, which resembles the introduction of the even factors during the system evolution discussed in the last paragraph. So, we argue that the model considering the even factors during the system evolution may be valid only for the C-C systems with a "wise manipulator", which are rarely observed in nature. This is why it is difficult to obtain empirical supports for the model considering both the Matthew effect and the even factors. This discussion supports that the current model should describe the common feature of most real world C-C systems.

In this work, although we tried our best to collect as many data as possible, the difficulties in collecting all the necessary information, especially the cooperation sharing and evolution duration, renders the collected data still far from sufficient. Apparently, collecting more data with all the necessary information will be highly useful to more robustly test the above proposed evolution model in the future work.

**Acknowledgements**

This work is supported by the National Natural Science Foundation of China under grant No. 10635040 and 70671089 and benefited of very helpful discussions with Dr. Shunguang Wu, Prof. Tao Zhou, Prof. Zong-Hua Liu and Prof. Hong Zhao.